\documentclass[12pt]{iopart}
\newcommand{\STO}{{LSMO$_{STO}\:$}}
\newcommand{\NGO}{{LSMO$_{NGO}\:$}}
\newcommand{\LSAT}{{LSMO$_{LSAT}\:$}}
\newcommand{\LAO}{{LSMO$_{LAO}\:$}}
\usepackage{graphicx}
\begin{document}

\title[Ab initio electronic structure in strained LSMO]{Ab-initio electronic and magnetic structure in La$_{0.66}$Sr$_{0.33}$MnO$_3$: strain and correlation effects}

\author{Chunlan Ma$^{1,2}$, Zhongqin Yang$^1$ and Silvia Picozzi$^3$}
\address{$^1$Surface Physics Laboratory (National Key Laboratory), Fudan University,
Shanghai, 200433, China \\
$^2$Department of Applied Physics, University of Science and
Technology of Suzhou, Suzhou, 215011, China \\
$^3$CNR-INFM, CASTI Regional Lab. c/o Dip.  Fisica, Universit\`a
degli Studi dell'Aquila, 67100 Coppito (L'Aquila),Italy\\}
\ead{\mailto{silvia.picozzi@aquila.infn.it}}
\begin{abstract}
The effects of tetragonal strain on electronic and magnetic
properties of strontium-doped lanthanum manganite,
La$_{2/3}$Sr$_{1/3}$MnO$_3$ (LSMO),  are investigated by means of
density-functional methods. As far as the structural properties are
concerned, the comparison between theory and experiments for LSMO
strained on the most commonly used substrates, shows  an overall
good agreement: the slight overestimate (at most of 1-1.5 $\%$) for
the equilibrium out-of-plane lattice constants points to possible
defects in real samples. The inclusion of a   Hubbard-like
contribution on the Mn $d$ states, according to  the so-called
``LSDA+U" approach, is rather ineffective from the structural point
of view, but much more important from the electronic and magnetic
point of view. In particular, full  half-metallicity, which is
missed  within a bare density-functional approach, is recovered
within  LSDA+U, in agreement with experiments. Moreover, the
half-metallic behavior, particularly relevant for spin-injection
purposes, is  independent on the chosen substrate and is achieved
for all the considered  in-plane lattice constants. More generally,
strain effects  are not seen to crucially affect the electronic
structure: within the considered tetragonalization range, the
minority gap is only slightly ({\em i.e.} by about 0.1-0.2 eV)
affected by a tensile or compressive strain. Nevertheless, we show
that the growth on a smaller in-plane lattice constant can stabilize
the out-of-plane vs in-plane $e_g$ orbital and significatively
change their relative occupancy. Since $e_g$ orbitals are key
quantities for the double-exchange mechanism, strain effects are
confirmed to be  crucial for the resulting magnetic coupling.
\end{abstract}

\vspace{2pc}
\maketitle

\section{Introduction}

Strontium-doped lanthanum manganites, La$_{1-x}$Sr$_x$MnO$_3$, have been heavily studied in the past few decades\cite{jvs,rmp,coey}, due to their fascinating properties and intriguing applications. Nevertheless, most of the properties are composition-dependent and the most exotic ones, such as colossal magneto-resistance and/or half-metallicity, appear only when 0.17$<x<$0.5\cite{rmp}. In the case of the optimized composition ($x\sim$1/3), La$_{0.66}$Sr$_{0.33}$MnO$_3$ (denoted as LSMO hereafter) shows a spin polarization $>$95$\%$\cite{bowen} and a high Curie ferromagnetic temperature ($T_C\sim$ 370 K); therefore, this compound appear as a promising candidate
in the spin injection framework, especially since it is envisaged that LSMO-based devices could operate at room temperature.

The interactions between electrons and lattice distortions are well
known to play a significant role in the physics of the compound:  Mn
$e_g$ electrons, which are responsible for the rich variety of
attractive properties shown by manganites, are coupled to the
lattice through Jahn-Teller effects \cite{millis}.  In particular,
MnO$_6$ octahedral tetragonal distortions can be induced by the
lattice mismatch between the film and the substrate, to which
magnetic anisotropy and magnetoresistance effects of LSMO films were
found to be extremely sensitive\cite{tsui}. In particular, the
magnetization easy axis was shown to move from in-plane to
out-of-plane upon moving from tensile to compressive strain
\cite{aniso,ranno}. Moreover, it  is mandatory for industrial
applications to achieve a good control over the film growth, which
in turn requires a careful  choice of appropriate substrates to get
optimized LSMO films. High-performance epitaxial manganites thin
films have been successfully grown on [001]-oriented oxides, such as
LaAlO$_3$ (LAO)\cite{tsui,lee},
La$_{0.3}$Sr$_{0.7}$Al$_{0.35}$Ta$_{0.35}$O$_9$ (LSAT)\cite{tsui},
NdGaO$_3$ (NGO)\cite{tsui}, and SrTiO$_3$ (STO)\cite{bowen,maurice,bertacco}.
Strained manganites epilayers often show a different behavior with
respect to the bulk. In particular, for LSMO, tensile strain was
shown to reduce the Curie ordering temperature, $T_C$, and this was
successfully explained in terms of double exchange: the increase of
the in-plane  Mn-O bond length leads to a reduction of the hopping
term between Mn$^{3+}$-Mn$^{4+}$, therefore reducing $T_C$ (see
below)\cite{millis}. However, for other manganites, such as
La$_{1-x}$Ba$_{x}$MnO$_3$, a tensile strain was found to enhance
$T_C$ \cite{zhang}. Therefore, a thorough understanding has not yet
been achieved, due to contradictory results. Possible reasons might
include the still poor control over the samples growth and
stoichiometry. Moreover, one should keep in mind that the delicate
interplay between charge, spin and orbital ordering can be very
easily altered in manganites and the subtleties of the relevant physical effects (such
as magnetoresistance and anisotropy) can lead to strikingly
different results induced by generally tiny changes.

Despite the large abundance of experimental studies devoted to
substrate-induced strain effects on the LSMO magnetic
properties\cite{tsui,lee} and {\em ab-initio} theoretical works
focused on LSMO,\cite{pickett,jarlborg,banach,zenia,geng} a deep
investigation of strain effects on the electronic and magnetic
properties from first-principles (both with and without additional
correlation effect beyond standard exchange-correlation functionals) is lacking and will
therefore be the focus of the present work. We choose four typical
substrates whose mismatches with LSMO films induce different types
of lattice strain: LAO, for which the film undergoes a relatively
large in-plane compressive strains; LSAT and NGO, for which the film
undergoes a very small compressive strain; STO, for which the film
experiences a small tensile strain. The paper is organized as
follows: in Sec.\ref{tech} we report some technicalities (in terms
of structural and computational details); in Sec.\ref{struct} we
focus on the structural equilibrium properties  of LSMO strained on
different substrates. Then, we move in Sec.\ref{properties} to the
discussion of the effects of strain and correlation on the LSMO
electronic and magnetic properties, in terms of density of states
and magnetic moments. Finally, we summarize our conclusions in
Sec.\ref{concl}.

\section{Crystal structure and computational details}
\label{tech}
Our calculations are performed following the generalized gradient approximation (GGA)\cite{pbe} to the exchange-correlation potential within the framework of density functional theory (DFT), using the all-electron full-potential linearized augmented plane-wave (FLAPW)\cite{flapw} method as implemented in the
FLEUR-code\cite{fleur}. In order to take into account correlation effects beyond the local spin density approximation (LSDA) or the GGA, we used the so called ``LSDA+U" (or ``GGA+U", in this case) scheme\cite{anisimov} in the ``atomic limit" approximation, as implemented in FLEUR\cite{shick}; this formalism includes a Hubbard--like potential
acting on the Mn 3$d$ states that enhances the tendency towards localization which erroneously  lacks in bare GGA.

The unit cell is divided into non-overlapping spheres and interstitial region.  The muffin-tin (MT) radii are set to 2.5 a.u. for Sr and La, 2.0 a.u. for Mn and 1.5 a.u. for O, respectively.
The wave function expansion in the interstitial region was carried up to {\bf k$_{max}$} = 3.8 a.u.$^{-1}$ which limits the number of basis functions to be 600 or so. We treated  La 5s, 5p; Sr 4s, 4p as local-orbital states\cite{lo},  whereas La 4$f$ were treated as valence states. The  energy position of these latter states (resulting to be at $\sim$2 eV above the
Fermi level, see below) can be questionable and might be an artifact of DFT-FLAPW; however, 
tests made applying a large Coulomb parameter (U=10 eV) on La 4$f$ states show that the effect on the physical quantities we are mainly interested in ({\em i.e.} total energies and density of occupied states) is negligible.

In order to sample the irreducible Brillouin zone, 72 {\bf k}-points were used, according to the 12x12x4 Monkhorst-Pack shell\cite{mp}; the convergency with respect to this parameter was accurately tested.
Within the GGA+U formalism, we considered different values of U (such as 2 and 3 eV), which were found to give the better agreement of DFT results
with recent photoemission measurements.\cite{arpes} The J value was set to J = 0.7 eV.
\begin{figure}
\begin{center}
\includegraphics[scale=0.4,angle=0]{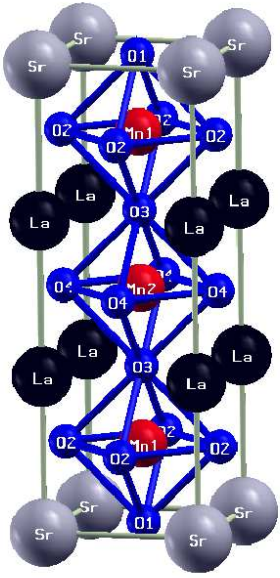}
\end{center}
\caption{Schematic unit cell (Bravais lattice shown by thin solid lines) for LSMO: grey, blue, red and black spheres denote Sr, O, Mn and La atoms, respectively. The vertical axis shows the [001] direction. O-O bonds are shown via  blue rods.}
\label{cell}
\end{figure}

In order to simulate LSMO, we used a 15-atoms unit cell which can be
schematically represented as a [001]-ordered
(SrMnO$_3$)$_1$/(LaMnO$_3$)$_2$ superlattice (see Fig.\ref{cell}),
similar to previous works\cite{pickett}. With this choice of
the unit cell, there are two inequivalent Mn: Mn$_1$ has a LaO plane
on one side and a SrO plane on the other side, whereas Mn$_2$ has
LaO layers on both parts.  Possible
 MnO$_6$ octahedral tiltings  were neglected. Effects
 induced by this approximation are quantitatively hard to predict; moreover, experiments
 suggest the tilting angle not to be strongly affected by different strain conditions.\cite{souza-neto-prb}
 The in-plane lattice constants $a$ are chosen according to the LAO,
NGO, LSAT and STO bulk experimental values\cite{tsui};  this is of course an approximation, and the theoretical substrate lattice constants might be used, as well. In some cases ({\em i.e.} when the disagreement between the theoretical and experimental lattice constants for the substrate or for the epilayer is rather high), one or the other choice can lead to different strain signs and qualitatively different behavior. However, in the present case, when considering LSMO grown on different substrates, the choice of the experimental substrate lattice 
constants does not change the strain sign with respect to the experimentally realized situation. In fact, the LSMO equilibrium calculated (see below) and experimental ``cubic" lattice constant are $a_{eq}^{th}$ = 3.90 Ang and $a_{eq}^{expt}$ = 3.874 Ang, respectively. Therefore, in both simulations and  experiments, the epilayer is under tensile strain for STO and under compressive strain in the other cases.
The out-of-plane lattice parameters $c$ have been set to {\em i})
experimental values for LSMO thin films\cite{tsui} grown on
different substrates and {\em ii}) minimized according to the
calculated total energies (see below). To test the effect of
structural optimization, we relaxed the atomic positions of LSMO
strained on STO until the forces were less than 0.002 Hartree/(a.u.). We
verified that the electronic structure is negligibly affected by
structural optimization of internal atomic coordinates; therefore, we will focus in the following
on the unrelaxed structures. In all the simulations, the three Mn
atoms in the unit cell were aligned ferromagnetically, according\cite{rmp} to
experiments\cite{notaCE}. 

\section{Structural equilibrium properties}
\label{struct}
\begin{figure}
\begin{center}
\includegraphics[scale=0.7]{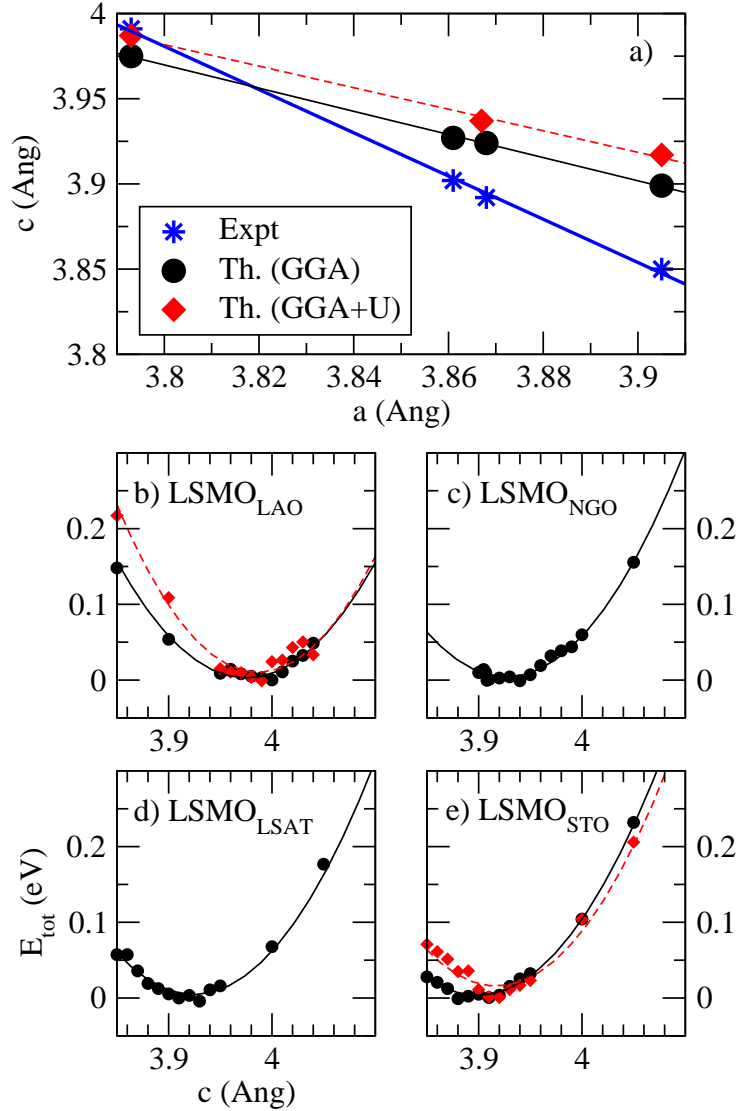}
\end{center}
\caption{a) Out-of-plane lattice constants (denoted as $c$) vs in-plane lattice constant (denoted as $a$): experimental values, GGA and GGA+U results are shown as blue stars, black filled circles and red filled diamonds, respectively. Linear fits to each set of data are also shown. The total energy curves as a function of $c$ are shown (with black symbols and solid lines) for different substrates in panels b), c), d) and e)
for \LAO, \NGO, \LSAT and \STO, respectively. In b) and e), the red symbols and dashed line show the results of GGA+U calculations (U=2 eV). Axes labels in panels b), c) and e) (not reported for clarity in plotting) are the same as d).}
\label{equil}
\end{figure}

In Fig.\ref{equil} we show the total-energy vs out-of-plane lattice constant for different substrates
(see panels b), c), d) and e) for \LAO, \NGO, \LSAT and \STO, respectively); we performed a parabolic fit
of the total energies and obtained the equilibrium $c$ as the parabola minimum for each different substrate. The resulting $c$ values are then plotted in Fig.\ref{equil} a)  as a function of the substrate
lattice constants and compared with corresponding experimental values. Similarly,
in Table \ref{tensicomp} we report the in-plane lattice constants ($a$) for the different chosen substrates,
as reported in Ref.\cite{tsui}, along with the corresponding experimental and theoretical values for the out-of-plane lattice constant ($c$), the strain components ($\epsilon_{ii}$), the $c/a$ ratio as well as the resulting volume ($V$).
The LSMO equilibrium ({\em i.e.} unstrained) lattice constant was estimated as the weighted average between the experimental lattice parameter of the constituents, LaMnO$_3$ and SrMnO$_3$ and is evaluated as $a_{eq}$ = 3.874 \AA.\cite{maurice}

First of all, as expected from elasticity theory, the $c$ parameter decreases as the strain becomes more and more tensile.
However, both from the theoretical and experimental point of view, the lack of a constant volume  shows that the material does not behave as an ideal elastic medium: the more compressive (tensile) is the strain, the  smaller (larger) is the volume.
Moreover, we see from Fig.\ref{equil} a) that
 in the considered range of substrate lattice constants, the $c$ vs $a$ relation is well described using a linear fit, both for theoretical as well as experimental values.
 However, the slope of the experimental results is larger than the theoretical one. In particular, GGA
slightly  overestimates the $c$ parameter in all cases (at most by
1.3$\%$), except for \LAO. This can be interpreted as follows: on
one hand, the slight disagreement between experimental and
theoretical $c$ values might be due to non-perfect experimental
samples. In fact, it is well known that twin-like defects\cite{maurice} or oxygen
lack - or excess - can often occur as a result of the different
growth conditions. The experimental films might therefore show a
non-perfect stoichoimetry or a non-high-quality crystallinity (at
variance with our simulated perfect crystal), therefore resulting in
slightly different lattice constants. On the other hand, we have to
keep in mind that GGA is known to describe rather well ({\em i.e.}
within 1-2 $\%$) the structural properties of magnetic systems
(generally better than LSDA), though showing, in some cases a slight
overestimation of the lattice constants. This could be the case,
although the disagreement with experiments is within the generally
acknowledged error range.  Therefore, in order to check whether the
error is due to the neglect of correlation effects, we repeated the
calculations for the LAO and STO substrates (see red diamonds and
dashed lines in Fig.\ref{equil} b) and e), respectively), using a
GGA+U approach with U=2 eV. In both cases, we found a slight
increase of the $c$ parameter, which brings LAO in excellent
agreement with experiments, but STO to an even worse value. Our
results are perfectly consistent with previous studies on other
magnetic oxides\cite{nicola}: the addition  of a Hubbard-like
potential to bare DFT generally leads to only slight changes in the
structural propertes (and in particular to a larger volume,
consistently with our results), though it is generally much more
``effective" from the electronic structure point of view, bringing
about a much closer agreement with spectroscopic properties, compared
to DFT (see below).

\begin{table}
\caption{\label{tensicomp} Experimental and theoretical structural parameters for LSMO films grown on different substrates:  in-plane lattice constants ($a$, in \AA),
out-of-plane lattice constants ($c$, in \AA), $c/a$ ratio and unit-cell volume  ($V$, in \AA$^3$). Experimental values are taken from Ref.\protect\cite{tsui}. In parenthesis, we report the in-plane and out-of-plane strain components  defined as $\epsilon_{xx} = (a-a_{eq})/a_{eq}$ and $\epsilon_{zz} = (c-a_{eq})/a_{eq}$, respectively.
Strains are evaluated with respect to the ideally unstrained experimental cubic lattice constant ($a_{eq} = c_{eq}$ = 3.874 \AA).}
\begin{indented}
\item\begin{tabular}{|l|c|cc|cc|cc|}
\br
Substrate & $a$ ($\epsilon_{xx}$) & \multicolumn{2}{c|}{$c$  ($\epsilon_{zz}$)} &  \multicolumn{2}{c|}{$c/a$} &  \multicolumn{2}{c|}{$V$}\\
& & Expt. & Th. & Expt. & Th. & Expt. & Th. \\ \hline
\mr
LAO & 3.793 (-2.1$\%$) & 3.991  (3.0$\%$) & 3.971  & 1.052 & 1.048 & 57.42 & 57.19 \\
NGO & 3.861 (-0.3$\%$) & 3.902 (0.7$\%$)& 3.927 &1.011 & 1.017 & 58.17 & 58.54 \\
LSAT & 3.868 (-0.1$\%$) & 3.892  (0.5$\%$)& 3.924 &1.006 & 1.014 & 58.23 & 58.71 \\
STO & 3.905 (0.8$\%$) & 3.850   (-0.6$\%$)& 3.899 & 0.986 & 0.999 & 58.71 & 59.47 \\
\br
\end{tabular}
\end{indented}
\end{table}

\section{Electronic and magnetic structure}
\label{properties}
Before  starting the discussion of strain-induced effects on the magnetic  and electronic properties, we briefly comment on one of the systems of interests, namely LSMO strained on LAO,  in terms of density of states and magnetic moments. Next, we will focus on the comparison between this same system and others under different strain conditions ({\em i.e.} different substrate lattice constants), according to the experimental equilibrium structure; finally, we will focus on the systems on a fixed in-plane lattice constant and study the effects of the $c$ elongation.

\subsection{Electronic and magnetic structure of LSMO strained on LAO: hybridization and correlation effects}

\begin{figure}
\includegraphics[scale=0.8]{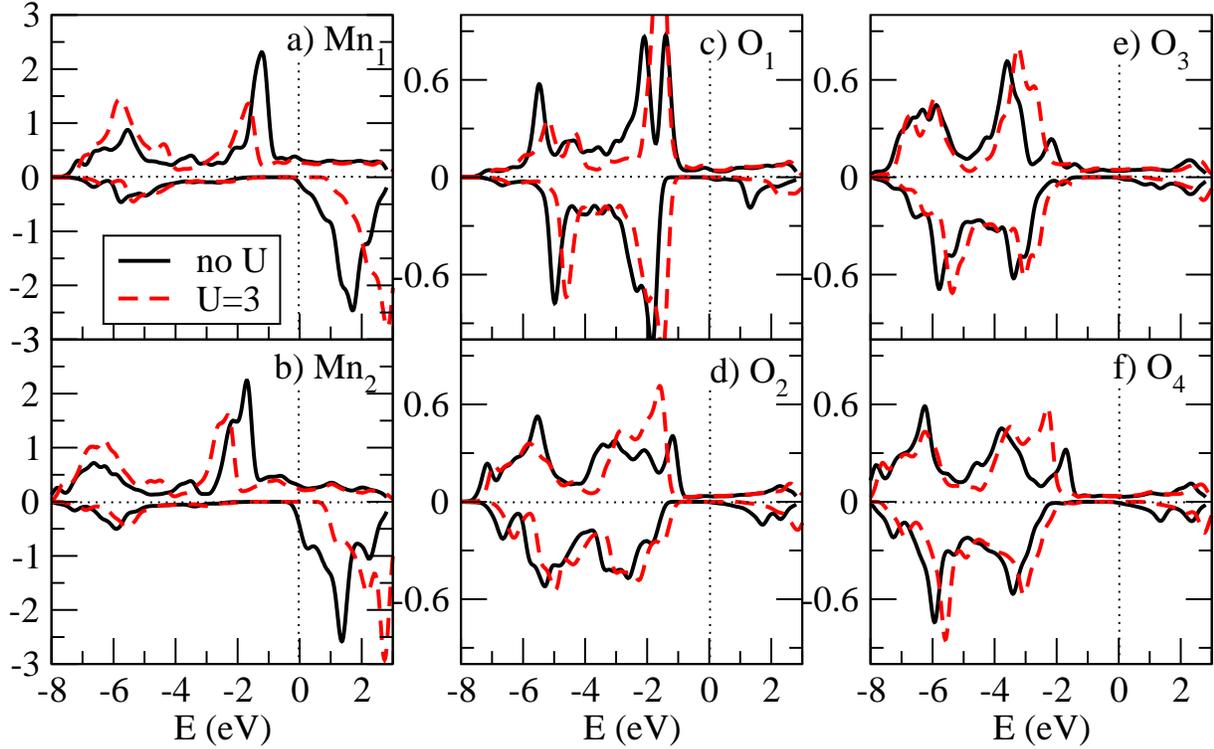}
\caption{PDOS for a) Mn$_1$, b) Mn$_2$, c) O$_1$, d) O$_2$, e) O$_3$ and f) O$_4$ for \LAO. Black solid and red dashed lines show the results of GGA and GGA+U calculations, respectively (U= 3 eV). Atomic labels are consistent with Fig.\protect\ref{cell}.}
\label{pdoslao}
\end{figure}

We show in Fig. \ref{pdoslao} the \LAO density of states projected
(PDOS) on the relevant atoms: Mn and O. As well known,  La and Sr
only donate their electrons to the electronic system, the bonding
being dictated essentially by the Mn-O interaction.\cite{pickett}.
In fact, as clearly visible in Fig.\ref{pdoslao}, there is a strong
hybridization between Mn $d$ and O $p$ states, especially in the
majority occupied states. The peak at binding energies at -1.2 eV
and -1.8 eV show the $t_{2g}$  orbitals (not so strongly
mixed), whereas the remaining part of the Mn PDOS shows the
$e_g$ contributions, much broader due to hybridization. As can be
also seen from the PDOS, and will be better clarified in the
following, the system is ``nearly half-metallic", {\em i.e.} it
shows a gap of $\sim$ 1-2 eV (see below), but the conduction band
minimum (CBM) is just  below  the Fermi level ($E_F$). Therefore,
the system is not strictly 100$\%$ spin-polarized at $E_F$.\cite{notaldaU4f} The
situation changes upon introduction of the Coulomb correlations: as
shown in Fig.\ref{pdoslao} in the case of U = 3 eV, the system
becomes fully half-metallic, consistently with experiments
suggesting a degree of spin--polarization larger than
95$\%$\cite{bowen}. As expected, the occupied (unoccupied) $d$
states are shifted towards higher binding (higher excited) energies
by about 1 eV ({\em i.e.} of the order of (U-J)/2) and, due to the
strong hybridization,  O $p$ states are also affected (see Fig.\ref{pdoslao} c), d), e) and f)).
Therefore, since within bare GGA half-metallicity is broken because
the tails of minority Mn $d$ states cross $E_F$, the Hubbard-induced
shift towards higher energies results in a fully half-metallic
character for \LAO.

As for the magnetic moments, we recall \cite{oxyvac,pickett} that
the two Mn atoms are surrounded by different cationic environments:
Mn$_1$ has a LaO plane on one side and a SrO plane on the other
side, whereas Mn$_2$ has LaO layers on both parts. As a result, the
magnetic moment of Mn$_2$ (3.26 $\mu_B$) is closer to pure LaMnO$_3$
in similar strain conditions ({\em i.e.} of the order of 3.3
$\mu_B$), whereas the moment of Mn$_1$ (3.11 $\mu_B$) is, as
expected, lower than Mn$_2$, but somewhat larger than the average of
LaMnO$_3$ and SrMnO$_3$ (this latter showing a muffin-tin moment  of about 2.7
$\mu_B$). The changes in the electronic structure induced by the
presence of additional correlations (and in particular the shift of
Mn $d$ states towards higher binding energies) affect  in turn the
magnetic moments: whereas, in the bare GGA case,  the total magnetic
moment is 10.89 $\mu_B$, this increases up to  11.0 $\mu_B$, when
the GGA+U approach is used. The integer total magnetic moment  is
consistent with the fully half-metallic character previously
outlined. Similarly, the atomic moments increase up to 3.29 $\mu_B$
and 3.49 $\mu_B$ for Mn$_1$ and Mn$_2$, respectively. However,
neither within GGA nor within GGA+U, the picture of charge ordering
and mixed Mn$^{3+}$ and Mn$^{4+}$ valencies is valid within the
present DFT-based formalism and with the present considered unit
cell, as already outlined in previous papers\cite{pickett}. Finally,
we remark that the magnetic moment is about 3.66 $\mu_B$/Mn, in
rather good agreement with experimentally reported values of 3.7
$\mu_B$ \cite{maurice}.

\subsection{Strain effects on the electronic and magnetic structure}

\begin{figure}
\begin{center}
\includegraphics[scale=0.7]{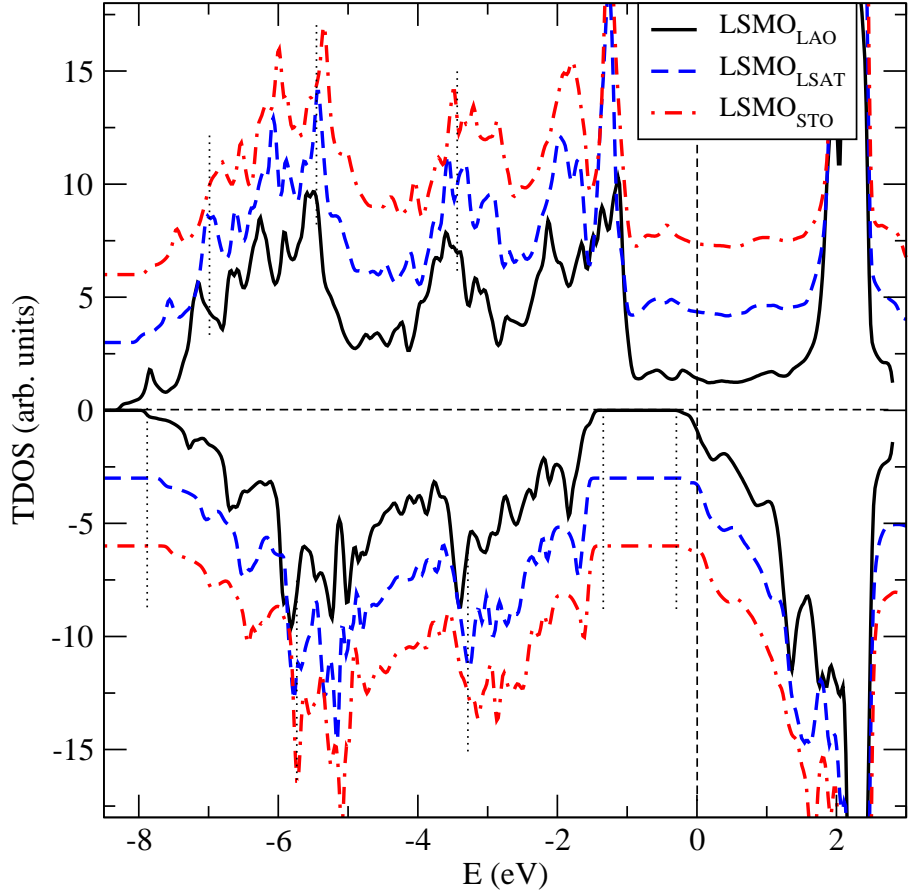}
\end{center}
\caption{TDOS for LSMO grown on LAO (black solid line), LSAT (blue
dashed line) and STO (red dot-dashed line)with GGA method. The zero
of the energy scale  is set to the Fermi level. Majority (minority)
spin contributions are shown in the positive (negative) $y$-axis.
For clarity, the TDOS in the case of \STO and \LSAT are arbitrarily
shifted along the $y$-axis. Vertical dotted lines are guide to the
eyes.} \label{tdos}
\end{figure}

Let us now focus on the structures grown on different substrates at the experimental $a$ and $c$ lattice constants (see Table \ref{tensicomp}). In Fig.\ref{tdos} we report the total density of states (TDOS) of the LSMO under tensile and compressive strains. The TDOS for LSMO grown on NGO looks very similar to LSMO grown on LSAT and is therefore not shown.

\begin{figure}
\begin{center}
\includegraphics[scale=0.8]{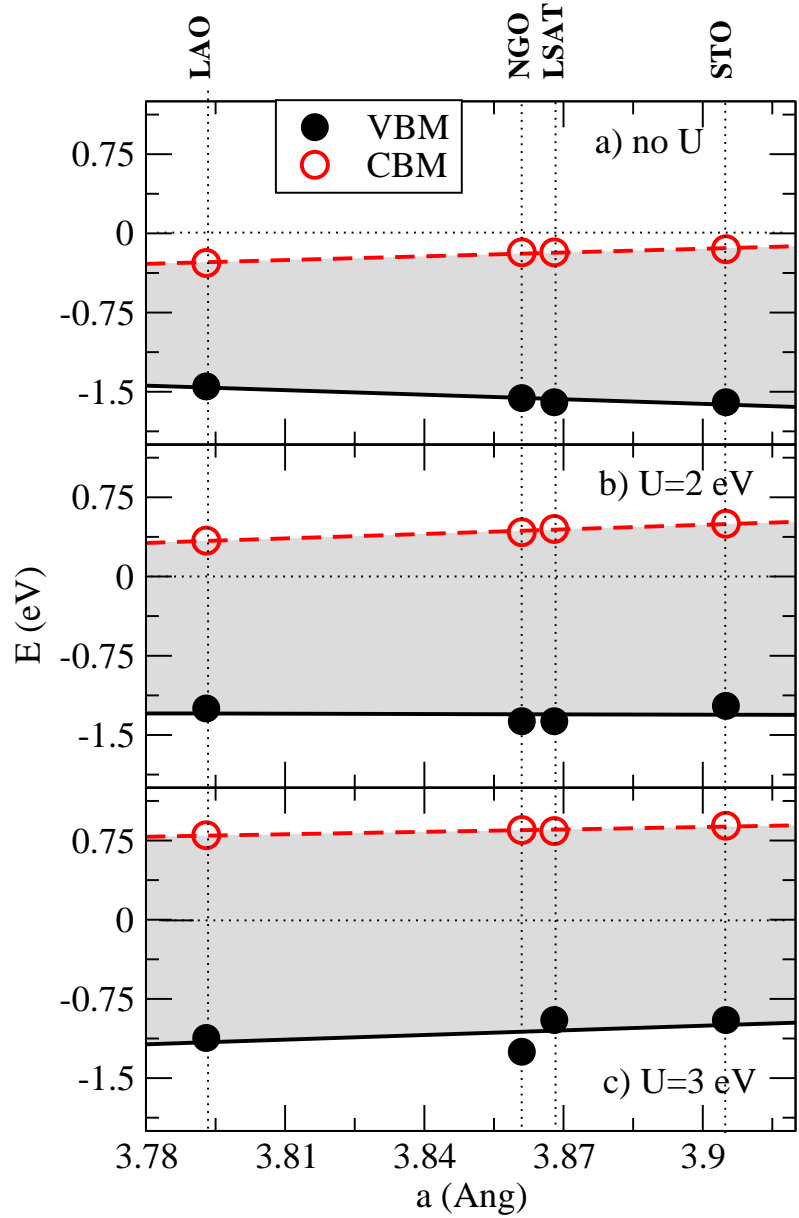}
\end{center}
\caption{CBM (red empty circles) and VBM (black filled circles)
positions in the minority spin channel with respect to $E_F$ (set to
zero) as a function of the in-plane lattice constant (labels on top
show the experimentally considered substrates). Panels a), b) and c)
show the results of bare GGA, GGA+U (U=2 eV) and GGA+U (U=3 eV)
approaches, respectively.} \label{vbmcbm}
\end{figure}

The comparison shows that the TDOS all look rather similar. However, one can note that, in going from
compressive to tensile strain, the band structure (both in the majority and minority spin channels) shifts towards lower binding energies (by $\sim$few tenths of eV for the considered strains of few percents). Furthermore,
in none of the considered substrates, the system shows full half-metallicity; rather, all of them are ``nearly" half-metallic. In order to further investigate the role of strain on the half-metallic behavior, we report in Fig.  \ref{vbmcbm} the energy position (taking $E_F$ as reference) of the valence band maximum (VBM) and of the CBM in the minority spin channel as a function of the substrate lattice constant.
The plots show that in all the considered systems, the CBM is always lower than $E_F$, therefore hindering the full half-metallicity. However, there is a slight tendency towards half-metallicity upon applying a tensile strain: LSMO grown on STO is closer to half-metallicity than LSMO grown on LAO. Furthermore,  the minority band-gap slightly increases with the substrate lattice constant.

Let us now focus on the effects of correlation and of strain on the band-gap and of the relative position of $E_F$ with respect to the minority band edges.
Upon introduction of U and already for a small value U = 2 eV, the Fermi level enters the gap. As expected, the gap (here defined as the difference between the CBM and the VBM, irrespective of the position of $E_F$) increases with applying progressively increasing U parameters: within the range of considered strains ({\em i.e.} in going from \LAO to \STO), it is of the order of 1.2-1.4 eV within bare GGA, of 1.6-1.8 eV within GGA+U (U = 2 eV) and of 1.9-2.1 eV within GGA+U (U = 3 eV).
On the other hand, no significative changes are observed upon applying strains (at least in the experimentally accessible range): the strain-induced effects on the gap-magnitude
amount at most to 0.1-0.2 eV.

It is of further interest to investigate the effects on the strain of the magnetic properties, in terms of magnetic moments.  In going from \LAO to \STO,
the Mn$_1$ and Mn$_2$
moments range from 3.11 $\mu_B$ to 3.15 $\mu_B$ and from 3.26 $\mu_B$ to 3.29 $\mu_B$, respectively. In parallel, the total moment ranges from 10.89 $\mu_B$ to 10.94 $\mu_B$. Therefore,
consistently with the slight changes in the electronic structure, the magnetic moments
only slightly increase upon increasing the substrate lattice constant.

Since the physics of LSMO crucially depends on the $e_g$ orbitals, whose occupancy and itineracy strongly affect the resulting ferromagnetism through double exchange,  we further investigate the Mn $d$ PDOS, further resolved by $m$ character. We recall that in a cubic symmetry, the $e_g$ level is twofold degenerate; however, upon tetragonal strain, the degeneracy is lifted and the $x^2-y^2$ in-plane orbital becomes different from the $3z^2-r^2$ out-of-plane orbital.
Therefore two competing mechanisms are induced by strain: {\em i}) on one hand, according to double exchange, $T_C$ is proportional to the transfer integral $t_0$, describing the Mn$^{3+}$-Mn$^{4+}$ hopping through intervening oxygen atoms. In turn, $t_0 \propto d_0^{-3.5}$ where $d_0$ is the in-plane Mn-O bond length. Therefore, a tensile strain, which increases the in-plane $d_0$, should reduce $t_0$, and, as a consequence, $T_C$; {\em ii}) on the other hand, the relative occupancy of the $x^2-y^2$ and
$3z^2-r^2$ orbitals also depend on strain: in particular, for a tensile strain, the occupancy of the in-plane orbital
should be higher than that of the out-of-plane orbital. In parallel, the $x^2-y^2$ has a larger transfer intensity: the stabilization and related increased occupation of this orbital therefore enhances electron hopping and, therefore, $T_C$.
The investigation of the strain-dependence of ferromagnetic coupling and of $T_C$
goes beyond the scope of the present work. However, we provide here some hints on the occupation of the $x^2-y^2$ and $3z^2-r^2$ orbitals, which could serve as a starting point for further investigations on the magnetism in LSMO under strain conditions.

\begin{figure}
\begin{center}
\includegraphics[scale=0.65]{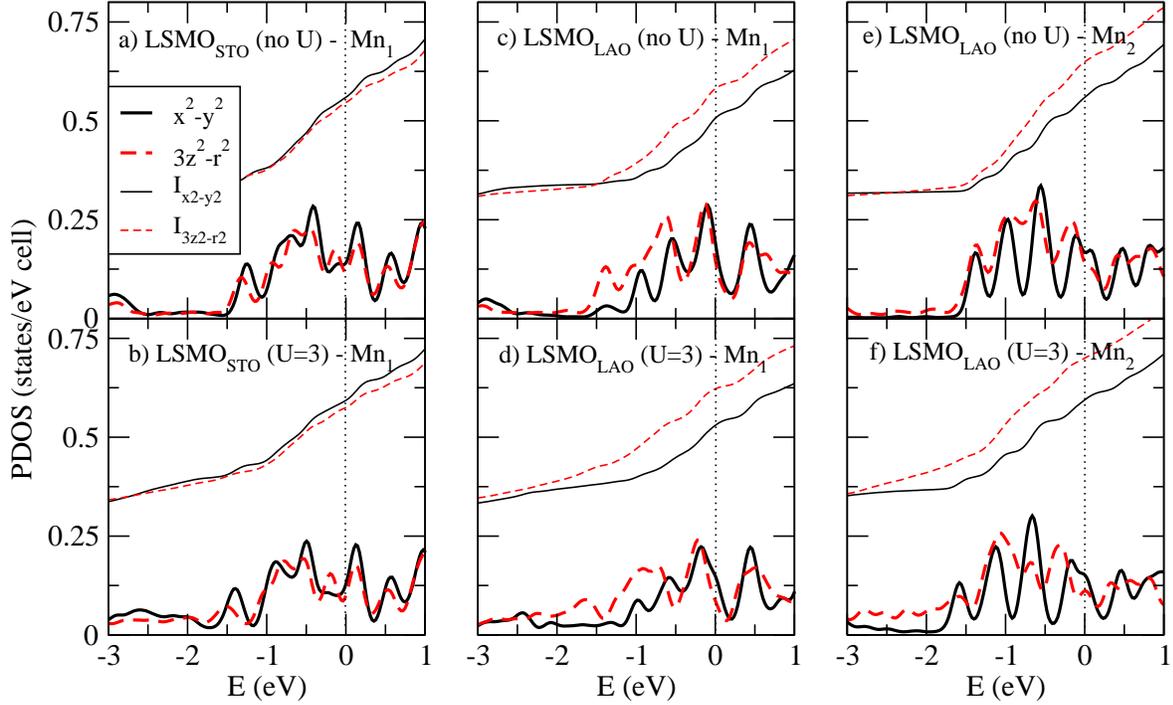}
\end{center}
\caption{Mn$_1$ PDOS of  $x^2-y^2$ (black bold solid) and $3z^2-r^2$ (red bold dashed) for LSMO on a) STO within
GGA, b) STO within GGA+U  (U=3 eV), c) LAO within GGA and d) LAO within GGA+U. The Mn$_2$ PDOS of  $x^2-y^2$ (black bold solid) and $3z^2-r^2$ (red bold dashed) for LSMO on e) LAO within GGA and f) LAO within GGA+U (U=3 eV). Also shown are the integrals: $I_{x^2-y^2}^{Mn}(E)$ (black thin solid)  and $I_{3z^2-r^2}^{Mn}(E)$ (red thin dashed).}
\label{dosorb}
\end{figure}

In Fig.\ref{dosorb}, we show the orbital-resolved
DOS projected on the two $e_g$ orbitals for \STO and \LAO (both with and without the inclusion of the Hubbard parameter U) for the two inequivalent Mn$_1$ and Mn$_2$ atoms.
We also show the integral of the PDOS, defined, for example for the in-plane $e_g$ orbital of Mn$_1$, as:
$I_{x^2-y^2}^{Mn_1}(E)=\int^E PDOS_{x^2-y^2}(E')\:dE'$. This is particular meaningful  for $E=E_F$, where the integral  coincides with the occupation of the orbital.
Note that the $t_{2g}$ orbitals as well are affected by strain and the three-fold degeneracy lifted. However, their
occupation is not modified upon tetragonal distortion: in fact, their PDOS (not reported) shows a peak at in the [-1.5:-1]-eV range and becomes negligible above $\sim$-0.5 eV,
resulting in full
occupation of the majority subband contributions. On the other hand, minority $t_{2g}$ states are completely unoccupied (as already mentioned, their tails slightly cross $E_F$ but this is likely to be an artifact of DFT and fully zero-occupation, {\em i.e.} half-metallicity,  can be restored by applying an additional Coulomb correlation).
It is evident from Fig.\ref{dosorb} a) that, even in LSMO$_{STO}$, showing an in-plane tensile strain of +0.8$\%$,
the degeneracy is lifted, but the consequences are still small: as expected, the in-plane $e_g$ is more occupied, but only by about 0.01 electrons more than in the out-of-plane $e_g$ case. This picture is kept more or less unaltered even upon application of U (see Fig. \ref{dosorb} b)).
On the other hand, if we look at \LAO,  where the strain is markedly compressive, there is a considerable difference between the occupations of the $e_g$ orbitals and the situation is reversed with respect to tensile strain: the $3z^2-r^2$ is about 15$\%$ more occupied than $x^2-y^2$ for Mn$_1$ within bare GGA (cfr Fig.\ref{dosorb} c)) and this same difference increases to 20$\%$ in the case of GGA+U
(cfr Fig.\ref{dosorb} d)) . Similarly and even more markedly, the occupation of the out-of-plane $e_g$ orbital is 20$\%$ higher for Mn$_2$ and further increases within GGA+U (cfr Fig.\ref{dosorb} e) and f)).
This is expected to have consequences on the ferromagnetism (although it is hard to give a quantitative estimate on the reduction of $T_C$ induced by the different occupations): in particular, upon compressive strain, the resulting lower transfer integral for the more-occupied $3z^2-r^2$ orbital should
suppress the in-plane carrier density and, as a consequence, double--exchange hopping and $T_C$, whereas, as already pointed out, a smaller in-plane MnO bond-length should lead to an opposite tendency.

\begin{figure}
\begin{center}
\includegraphics[scale=0.7]{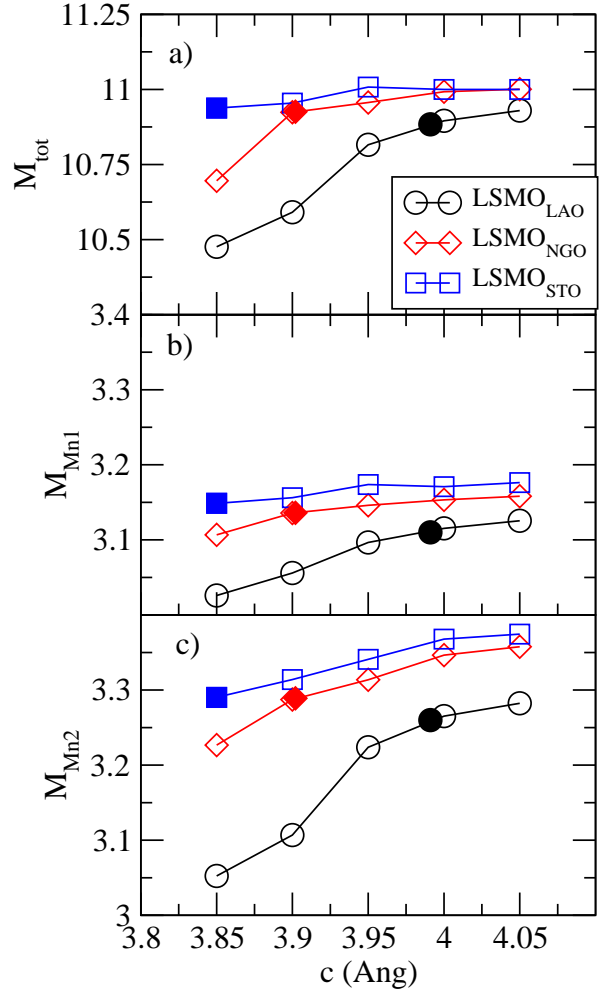}
\end{center}
\caption{a)  Total, b) Mn$_1$ and c) Mn$_2$ magnetic moments (in Bohr magnetons) as a function of the out-of-plane lattice constant (in \AA) for \LAO (black circles), \NGO (red diamond) and \STO (blue squares). Filled symbols show the values obtained for the experimental equilibrium structure.}
\label{momentc}
\end{figure}

Finally, we briefly comment on the effects of the Jahn-Teller $Q_3$-like distortions\cite{ahn} ({\em i.e.} octahedral elongation) for \LAO, \NGO and \LAO, in particular on the magnetic moments.
In Fig.\ref{momentc} we plot the magnetic moments (on the two Mn atoms and total moment per unit cell
in panels b), c) and a), respectively) as a function of the $c$ lattice constant for LSMO on the LAO, NGO and STO substrates, ;
we also show the experimental equilibrium structures using filled symbols.
As previously noted, the situation in \LSAT is very similar to \NGO and is therefore not shown.
It is worthwhile remarking that, whereas moments in \STO are basically unaltered by the out-of-plane lattice constants, \NGO shows  slightly larger strain-induced effects and in \LAO the
dependence on $c$ is rather significative. This is consistent with the related electronic structure: whereas LSMO on STO is always very close to half-metallicity for all the considered $c$ values, this condition is by far ``less satisfied" in LSMO grown on LAO. This has a relatively big effect on the magnetic moments: especially for lower $c$ values, there is an appreciable density of states coming from the minority conduction bands that cross $E_F$, therefore becoming occupied and lowering the total as well as Mn-projected magnetic moment.

\section{Conclusions}
\label{concl}

First-principles calculations have been performed on LSMO grown on different experimentally common substrates, in order to highlight the effects of uniaxial or biaxial strain on the electronic and magnetic structure. Our results show that:  {\em i}) GGA does not exactly reproduce the experimental structural parameters, such as the out-of-plane lattice constant,  and the
inclusion of correlation effects according the GGA+U formalism does not significatively alter
the situation. However, the discrepancy (at most of the order of 1.3 $\%$) falls within the generally acknowledged error range of density functional calculations and might also be due to defects in the experimental samples; {\em ii})  in none of the considered substrates, we obtain full half-metallicity within bare GGA. However, even the inclusion of a small Hubbard parameter U $\sim$ 2-3 eV - consistent with recent experiments - turns the system in a 100$\%$ spin-polarized density
of states at the Fermi level, in agreement with experiments and for all the considered substrates .
 Moreover, within bare DFT, \STO is closer to half-metallicity than \LAO: the minority $t_{2g}$ states, whose tails cross the Fermi level preventing full half-metallicity, have increasingly higher excited-state energies as the tensile strain grows. In summary, strain effects (at least in the experimentally accessible range) on the relevant electronic and magnetic properties, such as the half-metallic gap and the magnetic moments, do not significatively change the overall picture. However, subtle effects such as the occupation
 of the in-plane vs out-of-plane $e_g$ orbital,  that are quantitatively estimated in the case of \LAO and \STO, can change dramatically the magnetic coupling. These results call for further investigations on the strain-induced effects on the ferromagnetic exchange constants.

\ack{C.M. gratefully acknowledges kind hospitality from CNR-INFM CASTI regional lab, where 
part of this work was performed. We thank Prof. Dr. S. Bluegel and Dr. G.
Bihlmayer (Forschungszentrum Juelich, Germany) for interesting
discussions. The work is partly supported by the National
Natural Science Foundation of China under Grant No.10304002}


\begin{thebibliography}{10}
\bibitem{notaldaU4f} We observe that, if La 4$f$ are treated within LSDA+U (U = 10 eV),  changes
in the Mn $d$ unoccupied states around 2-3 eV are expected, but the half-metallic vs ``nearly" half-metallic character is unaffected.
\end{thebibliography}
\end{document}